\documentstyle[epsfig,12pt]{article}
\begin{document}

\title{A dynamical characterization of the small world phase}
\author{Tanya Ara\'{u}jo\thanks{%
Dept. Economia, UECE, ISEG, R. Miguel Lupi 20, 1200 Lisboa, Portugal
(tanya@iseg.utl.pt)}, R. Vilela Mendes\thanks{%
corresponding author} \thanks{%
Grupo de F\'{i}sica-Matem\'{a}tica, Complexo Interdisciplinar, Universidade
de Lisboa, Av. Gama Pinto 2, 1699 Lisboa Codex, Portugal
(vilela@cii.fc.ul.pt)}, Jo\~{a}o Seixas\thanks{%
Departamento de F\'{i}sica, Instituto Superior T\'{e}cnico, Av. Rovisco
Pais, 1096 Lisboa Codex, Portugal (seixas@fisica.ist.utl.pt)}}
\date{}
\maketitle

\begin{abstract}
Small-world (SW) networks have been identified in many different fields.
Topological coefficients like the clustering coefficient and the
characteristic path length have been used in the past for a qualitative
characterization of these networks. Here a dynamical approach is used to
characterize the small-world phenomenon. Using the $\beta -$model, a coupled
map dynamical system is defined on the network. Entrance to and exit from
the SW phase are related to the behavior of the ergodic invariants of the
dynamics.
\end{abstract}


\section{Introduction}

Networks are prevalent in all domains of life and science. Social, economic
and political networks are the backbone of human society. The internet is a
network. The metabolic processes of living beings form a network with the
substrates as nodes, which are linked whenever they participate in the same
biochemical reaction. Protein-protein and gene expression and regulation are
biological networks, etc.

Regular lattices and random graphs\cite{Erdos} have for a long time been
studied. More recently \cite{Watts1}-\cite{Watts2} small-world networks
became the object of growing attention and were identified in many different
fields. They seem to be the underling structure for some important phenomena
like the rapid spread of diseases, social networks, cooperative behavior
between competing agents, problem solving organization and communication
networks.

Topologically, small-world (SW) networks are identified by the values of two
statistical properties:

\begin{itemize}
\item  the {\it clustering coefficient} ({\it CC})\ that measures the
average probability for two agents, having a common neighbor, to be
themselves connected and

\item  the characteristic {\it path length }({\it PL}), this being the
average length of the shortest path connecting each pair of agents.
\end{itemize}

Regular lattices have long path lengths and high clustering, whereas random
graphs have short path lengths but low clustering. SW networks exhibit short 
{\it PL}'s and, at the same time, high {\it CC}'s.

In many model networks, the simultaneous occurrence of high {\it CC} and low 
{\it PL} is observed over an interval between order and randomness, which is
called the SW phase. However, this phenomenon can only be defined as a
phase, in the statistical mechanics sense, if order parameters are found to
characterize the regular-to-SW and the SW-to-random phase transitions.

Further information on the SW phenomenon has been obtained in the past from
the study of several quantities. Farkas et al. \cite{Farkas} studied the
spectral density of the adjacency matrix, with increasing randomness,
concluding that, in spite of the blurring of singularities, a consistently
high value of the third moment implies the existence of a large number of
triangles in the SW network. Monasson \cite{Monasson} on the other hand
studied the spectral properties of the Laplacian operator, that
characterizes the time evolution of a diffusive field, and localization
properties on the graph.

In this paper a dynamical systems approach is used to characterize the
small-world phenomenon. Using the $\beta -$model\cite{Watts2}, we study a
coupled map system on the network, with interactions defined by the network
connections. The SW phase is related to the behavior of the ergodic
invariants of the dynamics. Entrance to the SW phase is related to the
Lyapunov spectrum and exit from the SW phase corresponds to the region where
``entropy'' and ``conditional exponents entropy''\cite{Vilela1} \cite
{Vilela2} split apart.

\section{The dynamical model}

Consider a $\beta -$family of models, each one with $N$ agents on a circle
and periodic boundary conditions. For $\beta =0$, each agent in the model is
connected to its $2v$ nearest neighbors. For $\beta \neq 0$, the network
structure is obtained by looking at each one of the connections of the $%
\beta =0$ structure and, with probability $\beta $, replacing this
connection by a new random one.

On each one of the $\beta -$networks, a dynamical system is defined, with a
map at each node and convex-coupling interactions defined by the network
connections 
\begin{equation}
x_{i}(t+1)=\sum_{j=1}^{N}W_{ij}f\left( x_{j}(t)\right)   \label{2.3}
\end{equation}
where 
\begin{equation}
W_{ij}=\left\{ 
\begin{array}{lll}
1-\frac{n_{v}(i)}{2v}c & \textnormal{if} & i=j \\ 
\frac{c}{2v} & \textnormal{if} & i\neq j\textnormal{ and }i
\textnormal{ is connected to }j \\ 
0 &  & \textnormal{otherwise}
\end{array}
\right.   \label{2.4}
\end{equation}
$n_{v}\left( i\right) $ is the number of agents connected to $i$ and c is a
control parameter.

For the agents dynamics we choose 
\begin{equation}
f\left( x\right) =\alpha x\hspace{3.45cm}\textnormal{mod. } 1  \label{2.5}
\end{equation}
Typically $\alpha =2$.

For the $\beta =0$ network, each agent has exactly $2v$ neighbors and the
Lyapunov exponents are 
\begin{equation}
\lambda _{0}\left( k\right) =\log \left\{ \alpha \left( 1-c+\frac{c}{v}%
\sum_{j=1}^{v}\cos \left( j\theta _{k}\right) \right) \right\}  \label{2.6}
\end{equation}
with $\theta _{k}=\frac{2\pi k}{N}$ , $k=0,\cdots ,N-1$. In the $%
N\rightarrow \infty $ limit, the Lyapunov spectrum is a continuous smooth
function, as illustrated in the upper plot of Fig. 1. For reasons to be
discussed below we always choose $c$ in such a way that, for $\beta =0$, the
lowest Lyapunov exponent is zero.

\begin{figure}[htb]
\begin{center}
\psfig{figure=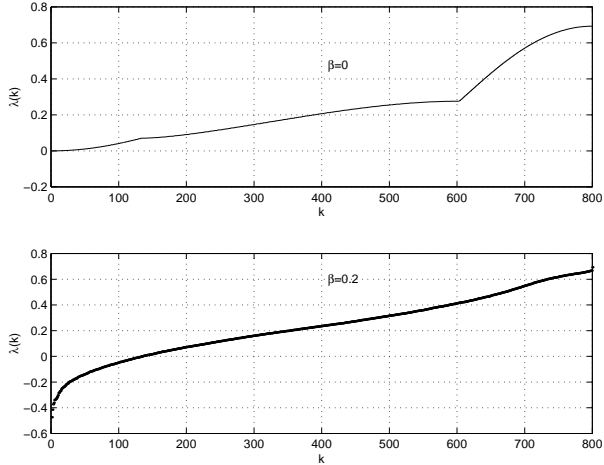,width=8truecm}
\end{center}
\caption{Lyapunov spectrum for $\beta =0$ and for a typical network at $%
\beta =0.2$ ($N=800$, $2v=6$)}
\end{figure}

As $\beta $ increases, the matrices of the tangent map cease to be regularly
organized, the Lyapunov spectrum develops gaps and some of the exponents
become negative. This is illustrated in the lower plot of Fig. 1 for $N=800$
and $2v=6$.

It is also the appearance of random long range connections that is
responsible for the reduction of the path length in SW networks. Therefore
it is natural to consider the modifications in the Lyapunov spectrum as the
dynamical signature of the onset of the SW phase. Of particular dynamical
significance is the shift of part of the spectrum towards negative values.
That is, the randomness arising from the rewiring leads to an effective
reduction of the dynamical degrees of freedom. We define $D_{\beta }$%
\begin{equation}
D_{\beta }=-\sum_{\lambda _{i}<0}\lambda _{i}  \label{2.7}
\end{equation}
to quantify this effect. To characterize the modifications of the Lyapunov
spectrum, another possibility would be to measure the singular part of the
spectrum associated to the gaps. However the natural intervals in the
spectrum associated to finite $N$ make this measurement less reliable.

In the upper plot of Fig. 2 we show the average values of $D_{\beta }$ taken
over $100$ different samples for each $\beta $ (with $N=800$ and $6$ as the
average degree of the network). A good fit to all the data shown in the
lower plot of Fig.2 is 
\begin{equation}
D_{\beta }=cN\left( \beta -\beta _{c_{1}}\right) ^{\eta _{1}}  \label{2.8}
\end{equation}
with $\beta _{c_{1}}\leq 10^{-5}$ and $\eta _{1}=1.01\pm 0.06$.

\begin{figure}[htb]
\begin{center}
\psfig{figure=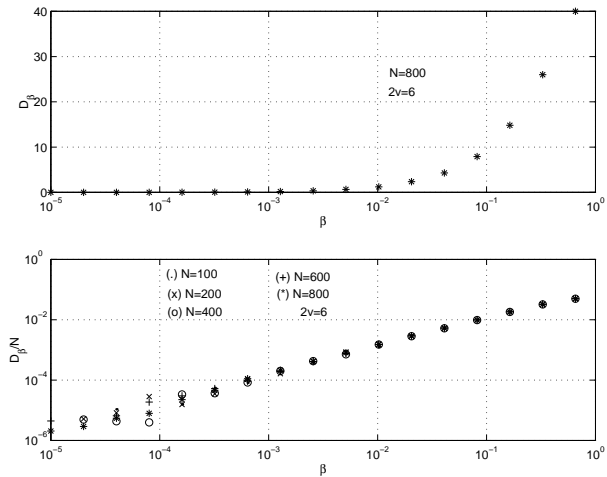,width=8truecm}
\end{center}
\caption{The $D_{\beta }$ parameter (``SW order parameter'') averaged over $%
100$ sample networks ($N=100,\cdots ,800$, $2v=6$)}
\end{figure}

In practice it is only after $\beta \simeq 10^{-3}$ that small-world effects
(and $D_{\beta }$ values) become appreciable. Nevertheless, the fact that
the data is consistent with $\beta _{c_{1}}=0$ implies that, using $\frac{1}{%
N}D_{\beta }$ as an order parameter for the small-world phase, this phase
starts at $\beta =0^{+}$, the regular phase being only the isolated point $%
\beta =0$.

To characterize the exit from the SW phase, we use the notion of {\it %
conditional Lyapunov exponents} . They were introduced by Pecora and Carroll
in their study of synchronization of chaotic systems\cite{Pecora}. Like the
Lyapunov exponents, the conditional exponents are well defined ergodic
invariants\cite{Vilela1}. The idea is that the conditions that in Oseledec's
theorem insure the existence of the Lyapunov exponents also establish the
existence of characteristic exponents formed by subblocks of the tangent map
matrix. Here, for each agent $i$, we consider a subblock of dimension $%
d_{i}\times d_{i}$ formed by himself and those that are connected to it. The
positive conditional exponents $\lambda _{\beta }^{*}(j)$ associated to each
subblock are computed and a dimension-weighed sum is performed over all
subblocks. This gives a version of what elsewhere\cite{Vilela1} \cite
{Vilela2} has been called a {\it conditional exponents entropy}. 
\begin{equation}
h_{\beta }^{*}=\sum_{i=1}^{N}\left( \frac{1}{d_{i}}\sum_{\lambda _{\beta
}^{*}>0}\lambda _{\beta }^{*}(j)\right)  \label{2.9}
\end{equation}
Subtracting $h_{\beta }^{*}$ from the sum of the positive Lyapunov
exponents, $h_{\beta }=\sum_{\lambda _{\beta }>0}\lambda _{\beta }(j)$, we
define the coefficient 
\begin{equation}
C_{\beta }=\left| \frac{h_{0}^{*}-h_{0}}{h_{\beta }^{*}-h_{\beta }}\right|
\label{2.10}
\end{equation}
which is also an ergodic invariant.

This coefficient has the following dynamical interpretation: The Lyapunov
exponents measure the rate of information production or, from an alternative
point of view, they define the dynamical freedom of the system, in the sense
that they control the amount of change that is needed today to have an
effect on the future. In this sense the larger a Lyapunov exponent is, the
freer the system is in that particular direction, because a very small
change in the present state will induce a large change in the future. The
conditional exponents have a similar interpretation concerning the dynamics
as seen from the point of view of each agent and his neighborhood \cite
{Vilela2}. However the actual information production rate is given by the
sum of the positive Lyapunov exponents, not by the sum of the conditional
exponents. Therefore, the quantity $h_{\beta }^{*}-h_{\beta }$ is a measure
of apparent dynamical freedom (or apparent rate of information production).
As self-organization in a system concerns the dynamical relation of the
whole to its parts, this quantity may also be looked at as a measure of
dynamical selforganization.

\begin{figure}[htb]
\begin{center}
\psfig{figure=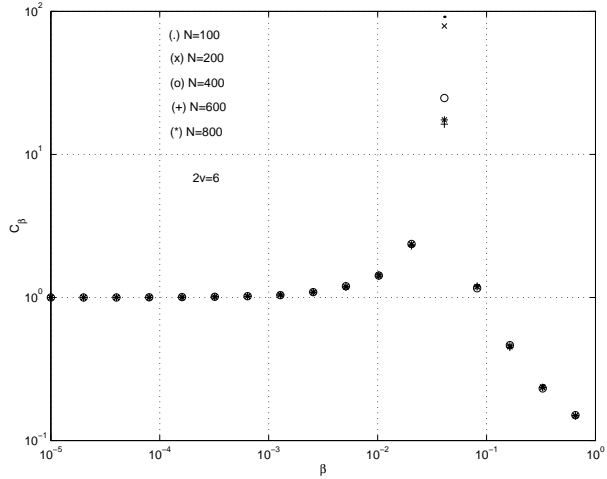,width=8truecm}
\end{center}
\caption{The $C_{\beta }$ parameter (``SW exit'') averaged over $100$ sample
networks ($N=100,\cdots ,800$, $2v=6$)}
\end{figure}

In Fig. 3 we show the average values of $C_{\beta }$ taken over $100$
different samples for each $\beta $ (with $6$ as the average degree of the
network and $N=100,200,400,600,800$). Notice the $N$-independence of $%
C_{\beta }$ which follows from the fact that, in Eq.(\ref{2.10}) it is
defined as a ratio of two quantities with the same $N$-dependence. For small 
$\beta $ values the difference between the entropy and the conditional
exponents entropy is a small quantity, that may be easily computed from the
network parameters. It means that each agent may have exact information on
the global behavior from observation of his own neighborhood. When $\beta $
increases the difference changes sign and becomes very large meaning that
the neighborhood information has ceased to provide reliable information on
the actual dynamics. This is the dynamical correlate of the decreasing
cluster properties and allows us to define the transition at the divergence
point $\beta _{c_{2}}$ of $C_{\beta }$. We find 
\begin{equation}
\beta _{c_{2}}\simeq 0.04  \label{2.11}
\end{equation}
Near the transition region 
\begin{equation}
C_{\beta }\sim \left| \beta -\beta _{c_{2}}\right| ^{-\eta _{2}}
\label{2.12}
\end{equation}
with $\eta _{2}\simeq 1.14$ below the transition and $\eta _{2}\simeq 0.93$
above it.

\section{Conclusions}

1 - The ergodic invariants (Lyapunov spectrum and conditional exponents)
provide a link between the topological properties of SW networks and the
dynamical behavior of a coupled map system modelled on the network. In
addition, the power laws obeyed by these invariants provide a framework to
identify the SW phenomenon as a phase in the statistical mechanics sense.

2 - Coupled map behavior, evolution of a diffusive field \cite{Monasson} and
spectrum of the adjacency matrix \cite{Farkas} supply complementary
information on the SW phenomenon. It is therefore conceivable that
quantities obtained from these other approaches might also be used to
construct order parameters characterizing the SW phase.

\end{document}